\documentclass[10pt]{article}
\usepackage[OE]{express}
\usepackage{graphicx}
\usepackage{amsmath}
\usepackage{amssymb}

\DeclareMathOperator{\sinc}{sinc}

\begin{document}

\title{General analytical solution for the electromagnetic grating diffraction problem}

\author{Alexandre V. Tishchenko,\authormark{1} and Alexey A. Shcherbakov,\authormark{2,*}}

\address{\authormark{1}Lab Hubert Curien, University of Lyon, 18 Rue Professeur Benoit Lauras, 42000 Saint-Etienne, France\\
\authormark{2}Laboratory of Nanooptics and Plasmonics, Moscow Institute of Physics and Technology, Institutsky 9, 141700 Dolgoprudnyi, Russia}

\email{\authormark{*}alex.shcherbakov@phystech.edu} %% email address is required

\begin{abstract}
Implementing the modal method in the electromagnetic grating diffraction problem delivered by the curvilinear coordinate transformation yields a general analytical solution to the 1D grating diffraction problem in a form of a $T$-matrix. Simultaneously it is shown that the validity of the Rayleigh expansion is defined by the validity of the modal expansion in a transformed medium delivered by the coordinate transformation.
\end{abstract}

{\small One print or electronic copy may be made for personal use only. Systematic reproduction and distribution, duplication of any material in this paper for a fee or for commercial purposes, or modifications of the content of this paper are prohibited.}

\url{https://www.osapublishing.org/oe/abstract.cfm?URI=oe-25-12-13435}


%%%%%%%%%%%%%%%%%%%%%%% References %%%%%%%%%%%%%%%%%%%%%%%%%

%%%%%%%%%%%%%%%%%%%%%%%%%%  body  %%%%%%%%%%%%%%%%%%%%%%%%%%
\section{Introduction}
The Rayleigh hypothesis (RH) was first formulated in \cite{Rayleigh1897} and then applied to the theory of diffraction gratings in \cite{Rayleigh1907}. If one considers the reflection of a plane wave from the plane interface between two homogeneous media, there exist only three waves: the incident plane wave, the reflected outgoing plane wave, and the transmitted (refracted) plane wave. Considering reflection from a sinusoidal interface, Rayleigh looked for a solution in a similar form, assuming that the field above and under the grating interface only consists of outgoing waves with constant amplitudes. Whereas such assumption is proved to be true for the regions outside the grating, it is considered as doubtful within the grating region by many researches, and any method based on the RH is still regarded as approximate (e.g., \cite{Edee2013,Nordam2013}).

Despite a number of works rationalizing a limited applicability of the RH \cite{Uretski1965,Petit1966,Millar1971,Pavageau1968,
Bates1975,Wirgin1980}, there was evidence calling into question the established theoretical limits \cite{Watanabe2004, Elfouhaily2006}. Furthermore, numerical validity of the RH for deep sinusoidal gratings, even for the correct near field simulation, which contradicted the admitted belief, was demonstrated in \cite{Tishchenko2009} (see also \cite{Wauer2009}). This article presents a theoretical analysis based on a concept formulated in \cite{Tishchenko2010}. The Chandezon Method (CM) \cite{Chandezon1980,Granet1998}, and the True Modal Method (TMM) \cite{Botten1981,Li1993,Tishchenko2005} reputed to be rigorous in the diffraction theory are used here to shed light on the problem. Both of these methods are well established and yield stable and correct results when applied to deep gratings. We show that the association of a basic CM idea (the coordinate transformation, which does not depend on any hypothesis) and the TMM technique (a construction of the modal basis of the true permittivity and permeability profile) leads to the demonstration of the validity of the RH providing that the modal expansion is complete, and to an analytical solution to the grating diffraction problem.

In the article we, first, show that the CM known to rigorously solve grating diffraction problems by means of a coordinate transformation is actually identical to the Rayleigh hypothesis provided the fields are represented in the basis of true modes of the transformed structure instead of a basis of the diffraction orders calculated by means of the Fourier decomposition of the transformed structure. Second, the coordinate transformation approach implemented on the basis of the modes of the transformed structure will be demonstrated to lead to an exact analytical solution to a wide range of grating problems. Moreover, such analytical solutions appear to be identical to those obtained directly on the basis of the RH. To our knowledge, this is the first time when the solution to a general diffraction problem is found in a closed analytical form. Two grating examples being important for practical applications (sinusoidal and saw-tooth profiles) are chosen to illustrate these two steps. At the end of the paper we discuss consequences of these results for electromagnetic simulation.

\section{Problem formulation and notations}
\begin{figure}[ht!]
\centering\includegraphics[width=8cm]{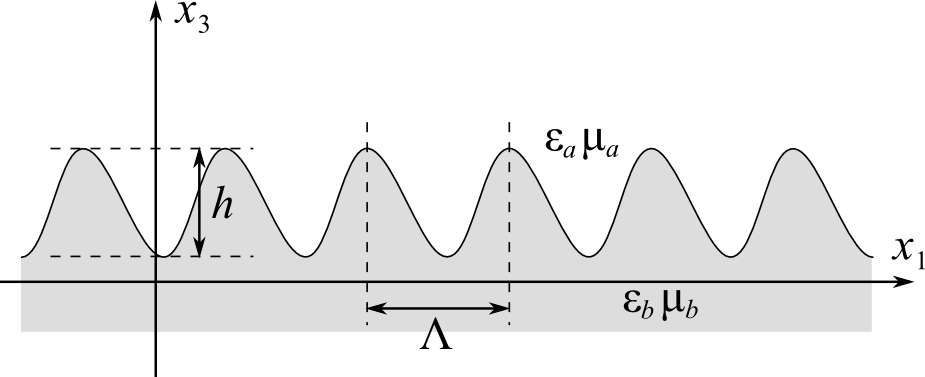}
\caption{Example of a grating corrugation.}
\label{fig:1}
\end{figure}

This work refers to the 1D plane grating linear diffraction problem which requires one to solve Maxwell's equations for a given incident field together with boundary conditions (continuity of tangential field components) at a periodically corrugated interface between two homogeneous isotropic media described by dielectric permittivities $\varepsilon_{a,b}$ and magnetic permeabilities $\mu_{a,b}$ (see illustration in Fig.~\ref{fig:1}). Due to linearity of the problem the electromagnetic fields will be implicitly assumed to be harmonic with $\exp(-j\omega t)$ time dependence factor.

Consider a Cartesian coordinate system $(x_1,x_2,x_3)$, whose axis $X_3$ is perpendicular to the grating plane, and axis $X_1$ indicates periodicity direction. Corrugation profile is supposed to be defined by a continuous and piecewise twice differentiable function $f(x_1)$ of period $\Lambda$, so that $f(x_1) = f(x_1+n\Lambda)$, $n\in\mathbb{Z}$.

According to the principles of the CM we will implement a transformation from Cartesian coordinates $(x_1,x_2,x_3)$ with unit orts $\bf{i}_{\alpha}$, $\alpha=1,2,3$, to curvilinear coordinates $(z^1,z^2,z^3)$, $z^{\alpha}=z^{\alpha}(x_1,x_2,x_3)$. Contravariant and covariant basis vector sets of the new coordinate system are $\bf{e}_{\alpha} = (\partial x_{\beta}/\partial z^{\alpha})\bf{i}_{\beta}$ and $\bf{e}^{\alpha} = (\partial z_{\alpha}/\partial x_{\beta})\bf{i}_{\beta}$ respectively. The two bases are mutually orthogonal: $\bf{e}_{\alpha}\cdot\bf{e}^{\beta} = \delta_{\alpha}^{\beta}$. Here and further summation over the repeating index is implied. Scalar products of basis vectors yield metric tensor components $g_{\alpha\beta} = \bf{e}_{\alpha}\cdot\bf{e}_{\beta}$ and $g^{\alpha\beta} = \bf{e}^{\alpha}\cdot\bf{e}^{\beta}$ with $g = \det\{ g_{\alpha\beta} \}$. Conventionally, covariant and contravariant components of any vector $\tilde{\bf{F}}$ are identified by lower and upper indices as $\tilde{F}_{\alpha}$ and $\tilde{F}^{\alpha}$, where the tilde is used to distinguish curvilinear vector components from components in the Cartesian coordinates $F_{\alpha}$. Corresponding relations read
\begin{equation}
	\begin{split}
		&F_{\alpha} = (\partial z^{\beta}/\partial x_{\alpha})\tilde{F}_{\beta}, \\
		&F_{\alpha} = (\partial x_{\alpha}/\partial z^{\beta})\tilde{F}^{\beta}.
	\end{split}
	\label{eq:field_transform}
\end{equation}
For more details on the tensor notations we refer readers to \cite{Schouten1954}.

Source-free Maxwell's equations in the Cartesian system read
\begin{equation}
	\begin{split}
	&\xi^{\alpha\beta\gamma}\frac{\partial E_{\gamma}}{\partial x_{\beta}} = i\omega\mu\delta^{\alpha\beta}H_{\beta}, \\
	&\xi^{\alpha\beta\gamma}\frac{\partial H_{\gamma}}{\partial x_{\beta}} = -i\omega\varepsilon\delta^{\alpha\beta}E_{\beta}.
	\end{split}
	\label{eq:Maxwell_X}
\end{equation}
Curls in left-hand sides of Eqs.~(\ref{eq:Maxwell_X}) are written via Levi-Civita symbols $\xi^{\alpha\beta\gamma}$, and Kronecker delta symbol $\delta^{\alpha\beta}$ is kept here for consistency with representation in curvilinear coordinates, where vector components with upper and lower indices differ. In curvilinear coordinates Maxwell's equations include the metric tensor components:
\begin{equation}
	\begin{split}
	&\xi^{\alpha\beta\gamma}\frac{\partial \tilde{E}_{\gamma}}{\partial z^{\beta}} = i\omega\mu\sqrt{g}g^{\alpha\beta}\tilde{H}_{\beta} \\
	&\xi^{\alpha\beta\gamma}\frac{\partial \tilde{H}_{\gamma}}{\partial z^{\beta}} = -i\omega\varepsilon\sqrt{g}g^{\alpha\beta}\tilde{E}_{\beta}.
	\end{split}
	\label{eq:Maxwell_Z}
\end{equation}

Eqs.~(\ref{eq:Maxwell_Z}) are quite similar to Eqs.~(\ref{eq:Maxwell_X}). The only difference is in the permittivity and the permeability tensors, which can be redefined as
\begin{equation}
	\tilde{\chi}^{\alpha\beta} = \chi\sqrt{g}g^{\alpha\beta}
	\label{eq:chi}
\end{equation}
with $\chi$ standing either for $\varepsilon$ or $\mu$. Such similarity leads to an important conclusion: a solution to the electromagnetic problem in the curvilinear coordinate system is equivalent to a solution in Cartesian coordinates $(\bar{x}_1,\bar{x}_2,\bar{x}_3)$ (bars are used here to avoid confusion with the initial coordinates) supposing that properties of the medium are determined in accordance with Eq.~(\ref{eq:chi}). We refer to this problem and the corresponding solution as reciprocal. This means that there exists a transformed medium with permittivity $\tilde{\varepsilon}$ and permeability $\tilde{\mu}$, in which the fields written in the Cartesian system $(\bar{x}_1,\bar{x}_2,\bar{x}_3)$ have the same coordinate dependences as the unknown fields $\tilde{\bf{E}}$ and $\tilde{\bf{H}}$ in the curvilinear coordinate system $(z^1,z^2,z^3)$. Every solution in the transformed medium corresponds to a solution of the initial problem providing that all the initial conditions and fields are correctly translated. The inverse transformation defined by Eqs.~(\ref{eq:field_transform}) yields the required solution in the initial Cartesian coordinates $(x^1,x^2,x^3)$. Note, that the transition from $(z^1,z^2,z^3)$ to global Cartesian coordinates $(\bar{x}_1,\bar{x}_2,\bar{x}_3)$ and back is not a coordinate transformation within a given diffraction problem, but a formal replacement of one diffraction problem with another, which is possible due to the similarity of Maxwell's equations (\ref{eq:Maxwell_X}) and (\ref{eq:Maxwell_Z}).

Thus, we will rely on the following claim. Given a boundary electromagnetic problem and a curvilinear coordinate system in which the boundary coincides with a coordinate plane, there exists a volume electromagnetic problem such as any solution in Cartesian coordinate system $(\bar{x}_1,\bar{x}_2,\bar{x}_3)$ has a corresponding solution to the initial problem in the transformed coordinate system $(z^1,z^2,z^3)$ expressed by the same coordinate functions.

\section{Coordinate transformation of the grating region}

The CM substitutes the electromagnetic grating diffraction problem by another one which deals with plane boundary but with changed permeability and permittivity tensors. The introduced tensors $\tilde{\varepsilon}$ and $\tilde{\mu}$ determine an electromagnetic response of the medium. Therefore, a choice of the curvilinear coordinate system is a very important step. If one chooses new coordinates so that tensor $g_{\alpha\beta}$ depends only on one coordinate, the reciprocal problem will be one-dimensional. This dramatically simplifies the resolution of the problem. In the considered case we can take
\begin{equation}
	\begin{split}
		&z^{1,2} = x_{1,2},\\
		&z^3 = z - f(x_1).
	\end{split}
	\label{eq:transform_zx}
\end{equation}
In accordance with Eqs.~(\ref{eq:chi}) and (\ref{eq:transform_zx}), the permittivity and permeability tensors of the transformed structure are
\begin{equation}
	\tilde{\chi}^{\alpha\beta} = \chi M^{\alpha\beta}
	\label{eq:chi_M}
\end{equation}
with
\begin{equation}
	M = \left( \begin{matrix} 1 & 0 & -f'(z^1) \\ 0 & 1 & 0 \\ -f'(z^1) & 0 & 1+[f'(z^1)]^2 \end{matrix}	 \right).
	\label{eq:M}
\end{equation}

\begin{figure}%[ht!]
	\centering\includegraphics[width=13cm]{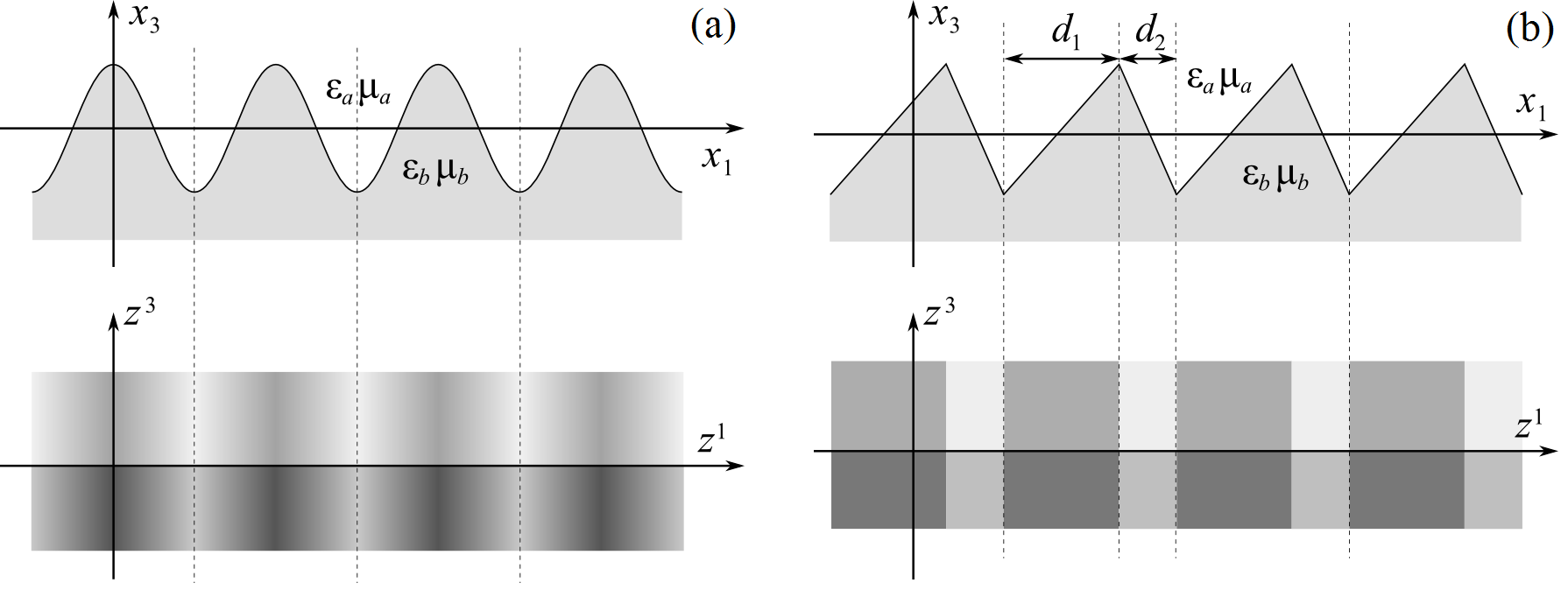}
	\caption{(a) an example of a sinusoidal corrugation separating two isotropic media, and a corresponding reciprocal problem in curvilinear coordinates described by smoothly varying material tensors. (b) an example of a saw-tooth corrugation separating two isotropic media, and a corresponding reciprocal problem in curvilinear coordinates described by two sets of the material tensors.}
  \label{fig:2}
\end{figure}

Consider here two illustrative examples. The first one is a sinusoidal corrugation between two isotropic media with profile function (the upper part of Fig.~\ref{fig:2}a):
\begin{equation}
	f(x_1) = \frac{h}{2}\sin\frac{2\pi x_1}{\Lambda}.
	\label{eq:sin-def}
\end{equation}
It follows from Eq.~(\ref{eq:M}) that
\begin{equation}
	M(z^1) = \left( \begin{matrix} 1 & 0 & -\frac{\pi h}{\Lambda}\cos\frac{2\pi z^1}{\Lambda} \\ 0 & 1 & 0 \\ -\frac{\pi h}{\Lambda}\cos\frac{2\pi z^1}{\Lambda} & 0 & 1+\left(\frac{\pi h}{\Lambda}\cos\frac{2\pi z^1}{\Lambda}\right)^2 \end{matrix}	 \right)
	\label{eq:M_sin}
\end{equation}
The permeability and the permittivity of the reciprocal medium appear to be smoothly modulated along the grating period, as the lower part of Fig.~\ref{fig:2}a illustrates.

The other practically important example is a saw-tooth corrugation (Fig.~\ref{fig:2}b):
\begin{equation}
	f(x_1) = \left\{ \begin{split} &\left(\frac{x_1}{d_1}-\frac{1}{2}\right)h,\;0\leq x_1<d_1; \\ &\left(\frac{\Lambda-x_1}{d_2}-\frac{1}{2}\right)h,\;d_1\leq x_1<\Lambda, \end{split} \right.
	\label{eq:saw-def}
\end{equation}
where $d_2 = \Lambda-d_1$. The coordinate transformation yields a reciprocal stratified structure described by two tensors per period
\begin{equation}
	\begin{split}
	&M_1 = \left( \begin{matrix} 1 & 0 & -h/d_1 \\ 0 & 1 & 0 \\ -h/d_1 & 0 & 1+(h/d_1)^2 \end{matrix}	 \right),\;0\leq x_1<d_1, \\
	&M_2 = \left( \begin{matrix} 1 & 0 & h/d_2 \\ 0 & 1 & 0 \\ h/d_2 & 0 & 1+(h/d_2)^2 \end{matrix}	 \right),\;d_1\leq x_1<\Lambda.
	\end{split}
	\label{eq:M_saw}
\end{equation}

\section{Modal solution to the diffraction problem}
The next step in the analysis is the resolution of the diffraction problem in the transformed medium with material tensors given by Eqs.~(\ref{eq:chi_M}) and (\ref{eq:M}) (with coordinates $(z^1,z^2,z^3)$ being substituted by $(\bar{x}_1,\bar{x}_2,\bar{x}_3)$). The transformed structure is composed of periodically stratified media being translational invariant along $\bar{x}_3$, in which an exact field solution (which also will be denoted with a bar as $\bar{F}$) can be expressed in terms of modes propagating up and down along coordinate $\bar{x}_3$. Once the modal basis of such volume grating is defined, any field solution in the grating region is represented by a superposition of grating modes \cite{Botten1981}. Since no dependence assumed along $\bar{x}_2$ direction, all grating modes split into TE and TM ones. The electric field of each TE mode is directed along the $\bar{X}_2$ axis, and so does the magnetic field of all TM modes.

Suppose that each mode propagates up or down the grating with the propagation constant $\beta_q$, $q\in\mathbb{Z}$. Since the grating interface corresponds to plane $\bar{x}_3 = 0$ in the reciprocal problem, the modal spectrum should be retrieved for tensors $\tilde{\varepsilon} = \tilde{\varepsilon}_a$ and $\tilde{\mu} = \tilde{\mu}_a$ in the region $\bar{x}_3 > 0$, and for $\tilde{\varepsilon} = \tilde{\varepsilon}_b$ and $\tilde{\mu} = \tilde{\mu}_b$ in the region $\bar{x}_3 < 0$. In other words the modal propagation constants and amplitudes will be different below and above the plane interface. Denote the propagation constant and the mode amplitude as $\beta_q^a$ and $a_q^{\pm}$, respectively, in the region $\bar{x}_3 > 0$. Here ``plus'' sign corresponds to the upward propagation. Analogously, the symbols $\beta_q^b$ and $b_q^{\pm}$ are used further to describe a mode of order $q$ in region $\bar{x}_3 < 0$.

Without loss of generality let us consider region the $\bar{x}_3 < 0$ (derivation for $\bar{x}_3 > 0$ is absolutely the same). The $\bar{x}_2$ component of the electric field of the $q$-th TE mode is
\begin{equation}
	\bar{E}^{TE}_{2q} = b_q^{e\pm}\exp\left[j\psi_q^b(\bar{x}_1) \pm j\beta_q^b\bar{x}_3\right],
	\label{eq:TE_E2}
\end{equation}
where function $\psi_q^b(\bar{x}_1)$ describes modal field distribution along the grating period. Field $\tilde{E}^{TE}_{2q}$ satisfies a quasi-periodicity condition:
\begin{equation}
	\bar{E}^{TE}_{2q}(\bar{x}_1+\Lambda) = \bar{E}^{TE}_{2q}(\bar{x}_1)\exp(jk^{inc}_{1}\Lambda)
	\label{eq:quasiperiodicity}
\end{equation}
imposed by an incident field with wavevector projection on $\bar{x}_1$ direction equal to $k^{inc}_{1}$. Substitution of Eq.~(\ref{eq:TE_E2}) into the first set of Maxwell's equations yields the magnetic field of the TE mode
\begin{equation}
	\begin{split}
	&\bar{H}^{TE}_{1q} = \frac{1}{\omega\mu_b} \left[\mp\beta_q^b \left\{1+[f'(\bar{x}_1)]^2\right\} + f'(\bar{x}_1)\frac{d\psi_q^b(\bar{x}_1)}{d\bar{x}_1} \right] b_q^{e\pm}\exp\left[j\psi_q^b(\bar{x}_1) \pm j\beta_q^b\bar{x}_3\right], \\
	&\bar{H}^{TE}_{3q} = \frac{1}{\omega\mu_b} \left[\mp\beta_q^b f'(\bar{x}_1) + \frac{d\psi_q^b(\bar{x}_1)}{d\bar{x}_1} \right] b_q^{e\pm}\exp\left[j\psi_q^b(\bar{x}_1) \pm j\beta_q^b\bar{x}_1\right].
	\end{split}
	\label{eq:TE_H13}
\end{equation}
Further substitution of this field components into the second set of Maxwell's equations provides a differential equation on function $\psi_q^b(\bar{x}_1)$:
\begin{equation}
	j\frac{d}{d\bar{x}_1}\left[ \pm\beta_q^bf'(\bar{x}_1)-\frac{d\psi_q^b(\bar{x}_1)}{d\bar{x}_1} \right] + \left[ \pm\beta_q^bf'(\bar{x}_1)-\frac{d\psi_q^b(\bar{x}_1)}{d\bar{x}_1} \right]^2 = \omega^2\varepsilon_b\mu_b - (\beta_q^b)^2,
	\label{mode_eq}
\end{equation}
which can be reduced to the Riccati type equation. Analogous considerations and derivations for the $q$-th TM mode yield the field components
\begin{equation}
	\begin{split}
	&\bar{H}^{TM}_{2q} = b_q^{h\pm}\exp\left[j\phi_q^b(\bar{x}_1) \pm j\beta_q^b\bar{x}_3\right],\\
	&\bar{E}^{TM}_{1q} = \frac{1}{\omega\varepsilon_b} \left[\pm\beta_q^b \left\{1+[f'(\bar{x}_1)]^2\right\} - f'(\bar{x}_1)\frac{d\phi_q^b(\bar{x}_1)}{d\bar{x}_1} \right] b_q^{e\pm}\exp\left[j\phi_q^b(\bar{x}_1) \pm j\beta_q^b\bar{x}_3\right], \\
	&\bar{E}^{TM}_{3q} = \frac{1}{\omega\varepsilon_b} \left[\pm\beta_q^b f'(\bar{x}_1) - \frac{d\phi_q^b(\bar{x}_1)}{d\bar{x}_1} \right] b_q^{e\pm}\exp\left[j\phi_q^b(\bar{x}_1) \pm j\beta_q^b\bar{x}_3\right]
	\end{split}
	\label{eq:TM_EH}
\end{equation}
and absolutely the same differential equation on function $\phi_q^b(\bar{x}_1)$ as Eq.~(\ref{mode_eq}). This means that all TE and TM modes of the same index have similar modal field distribution defined by solution $\psi_q^b(\bar{x}_1)\equiv\phi_q^b(\bar{x}_1)$ of Eq.~(\ref{mode_eq}).

General solution of Eq.~(\ref{mode_eq}) can be searched in form \cite{Murphy1960}
\begin{equation}
	\psi_q^b(\bar{x}_1) = \pm\beta_q^b f(\bar{x}_1) - j\log G(\bar{x}_1)
	\label{eq:psi-G}
\end{equation}
with an unknown function $G(\bar{x}_1)$. This function satisfies the homogeneous Helmholtz equation
\begin{equation}
	G''(\bar{x}_1) + \left[ \omega^2\varepsilon_b\mu_b - (\beta_q^b)^2 \right]G(\bar{x}_1) = 0.
	\label{eq:eqG}
\end{equation}
Therefore, in any region $R_i=\{\bar{x}_1:\bar{x}_{1,i-1}<\bar{x}_1<\bar{x}_{1,i}\}$, with $\bar{x}_{1,i}$ being some constants, a of continuous derivative $f'(\bar{x}_1)$ a general solution of Eq.~(\ref{mode_eq}) writes:
\begin{equation}
	\begin{split}
	&\psi_q^b(\bar{x}_1) = \pm\beta_q^b f(\bar{x}_1) \\&- j\log \left\{ C_i \exp\left[ j\sqrt{\omega^2\varepsilon_b\mu_b - (\beta_q^b)^2}\bar{x}_1 \right] + D_i \exp\left[-j\sqrt{\omega^2\varepsilon_b\mu_b - (\beta_q^b)^2}\bar{x}_1\right] \right\}
	\end{split}
	\label{eq:modal_solution}
\end{equation}
where constants $C_i$ and $D_i$ should to be related by vertical boundary conditions. Consider the modal fields at the vertical interface $\bar{x}_1 = \bar{x}_{1,i}$ between some two adjacent domains $R_i$ and $R_{i+1}$. Continuity of the tangent field component gives a remarkable result:
\begin{equation}
	\begin{split}
		&C_i = C_{i+i},\\&D_i = D_{i+1}.
	\end{split}
	\label{eq:CD}
\end{equation}
This means that each mode is either a pure right ($D_i=0$) or a pure left ($C_i=0$) propagating wave.

According to Eqs.~(\ref{eq:TE_E2}), (\ref{eq:quasiperiodicity}), and (\ref{eq:modal_solution}), dispersion equations for the right and left propagating modes are
\begin{equation}
		\exp\left[\pm j\Lambda\sqrt{\omega^2\varepsilon_b\mu_b - (\beta_q^b)^2}\right] = \exp(jk^{inc}_1\Lambda).
	\label{eq:dispersion}
\end{equation}
Thus, the propagation constant of the $q$-th order grating mode is
\begin{equation}
	\beta_q^b = \sqrt{\omega^2\varepsilon_b\mu_b - k^2_{1q}},
	\label{eq:beta}
\end{equation}
where $k_{1q} = k_1^{inc} + 2\pi q/\Lambda$, index $q$ runs from $-\infty$ to $\infty$, and $0\leq\arg(\beta_q^b)<\pi$. Additionally, substitution of the last relation into Eqs.~(\ref{eq:TE_E2}), (\ref{eq:TE_H13}), and (\ref{eq:TM_EH}) demonstrates that the absence of reflections at the vertical interfaces $\bar{x}_1 = \bar{x}_{1,i}$ is due to equality of mode impedances in all domains $R_i$:
\begin{equation}
	\frac{\bar{F}_{2q}}{\bar{G}_{3q}} = \frac{\omega\chi_b}{k_{1q}},
	\label{eq:impedance}
\end{equation}
where $F$ and $G$ stand for fields $E$ and $H$ for the TE polarization and vice versa for the TM polarization.

\section{Grating T-matrix}
In accordance with Eq.~(\ref{eq:beta}) the transverse modal field components relative to $\bar{x}_3$ direction are proportional to
\begin{equation}
	\begin{split}
	&\bar{F}_{2q} \sim b_q^{\pm}\exp\left\{ jk_{1q}\bar{x}_1 \pm j\beta_q^b[\bar{x}_3+f(\bar{x}_1)] \right\},\\
	&\bar{G}_{1q} \sim \frac{k_{1q}f'(\bar{x}_1)\mp\beta_q^b}{\omega\chi_b} b_q^{\pm}\exp\left\{ jk_{1q}\bar{x}_1 \pm j\beta_q^b[\bar{x}_3+f(\bar{x}_1)] \right\},
	\end{split}
	\label{eq:transverse}
\end{equation}
where notations are the same as for Eq.~(\ref{eq:impedance}), and $\chi$ denotes either $\mu$ or $\varepsilon$ in the TE or the TM case respectively, as before. Similar relations hold for the upper medium.

Continuity of the tangent field components at the interface $\bar{x}_3 = 0$ leads to the equations relating modal amplitudes below and above the interface:
\begin{equation}
	\begin{split}
		\sum\limits_{q=-\infty}^{\infty} & \left\{ b_q^+\exp\left[jk_{1q}\bar{x}_1 + j\beta_q^bf(\bar{x}_1)\right] + b_q^-\exp\left[jk_{1q}\bar{x}_1 - j\beta_q^bf(\bar{x}_1)\right] \right\} \\
		&= \sum\limits_{p=-\infty}^{\infty} \left\{ a_p^+\exp\left[jk_{1p}\bar{x}_1 + j\beta_p^af(\bar{x}_1)\right] + a_p^-\exp\left[jk_{1p}\bar{x}_1 - j\beta_p^af(\bar{x}_1)\right] \right\}.
	\end{split}
	\label{eq:continuity2}
\end{equation}
\begin{equation}
	\begin{split}
		\frac{1}{\chi_b}\sum\limits_{q=-\infty}^{\infty} & \left\{ \begin{matrix}[k_{1q}f'(\bar{x}_1)-\beta_q^b]b_q^+\exp\left[jk_{1q}\bar{x}_1 + j\beta_q^bf(\bar{x}_1)\right] \\+ [k_{1q}f'(\bar{x}_1)+\beta_q^b]b_q^-\exp\left[jk_{1q}\bar{x}_1 - j\beta_q^bf(\bar{x}_1)\right] \end{matrix} \right\} \\
		&= \frac{1}{\chi_a}\sum\limits_{p=-\infty}^{\infty} \left\{ \begin{matrix} [k_{1p}f'(\bar{x}_1)-\beta_p^a]a_p^+\exp\left[jk_{1p}\bar{x}_1 + j\beta_p^af(\bar{x}_1)\right] \\+ [k_{1p}f'(\bar{x}_1)+\beta_p^a]a_p^-\exp\left[jk_{1p}\bar{x}_1 - j\beta_p^af(\bar{x}_1)\right] \end{matrix} \right\}.
	\end{split}
	\label{eq:continuity1}
\end{equation}
It is known in the modal method theory that normalizing modal fields makes available analytical expressions for $T$-matrix components. To this end, we will use here the following integrals
\begin{equation}
	\begin{split}
		&\frac{1}{\Lambda}\int\limits_0^{\Lambda}\left[(\beta_p^{a,b}+\beta_q^{a,b})\mp(k_{1p}+k_{1q})f'(\bar{x}_1)\right] \exp\left[j(k_{1q}-k_{1p})\bar{x}_1\pm(\beta_q^{a,b}-\beta_p^{a,b}) f(\bar{x}_1)\right] d\bar{x}_1 \\
		& = 2\beta_p^{a,b}\delta_{pq},
	\end{split}
	\label{eq:normint1}
\end{equation}
\begin{equation}
	\begin{split}
		&\frac{1}{\Lambda}\int\limits_0^{\Lambda}\left[(\beta_p^{a,b}-\beta_q^{a,b})\pm(k_{1p}+k_{1q})f'(\bar{x}_1)\right] \exp\left[j(k_{1q}-k_{1p})\bar{x}_1\pm(\beta_q^{a,b}+\beta_p^{a,b}) f(\bar{x}_1)\right] d\bar{x}_1 \\
		&= 0,
	\end{split}
	\label{eq:normint2}
\end{equation}
which are proved in Appendix. Thus, multiplying Eqs.~(\ref{eq:continuity2}) by $[\beta_p^b\mp k_{1p}f'(\bar{x}_1)]\exp[jk_{1p}\bar{x}_1\mp j\beta_p^bf(\bar{x}_1)]/(2\beta_p^b\Lambda)$, and Eqs.~(\ref{eq:continuity1}) -- by $\mp\omega\chi_b\exp[jk_{1p}\bar{x}_1\mp j\beta_p^bf(\bar{x}_1)]/(2\beta_p^b\Lambda)$, combining them, integrating over the grating period, and applying orthogonality conditions of Eqs.~(\ref{eq:normint1}) and (\ref{eq:normint2}), one gets
\begin{equation}
	\begin{split}
		b_p^{\pm} = &\sum\limits_{q=-\infty}^{\infty} a_q^+ \frac{1}{\Lambda} \int\limits_0^{\Lambda} \left\{\begin{matrix} \frac{\left(\beta_p^b\pm \frac{\chi_b}{\chi_a}\beta_q^a\right)\mp\left(\frac{\chi_b}{\chi_a}k_{1q}+k_{1p}\right)f'(\bar{x}_1)}{2\beta_p^b} \\ \times\exp\left[j\Delta k_{1qp}\bar{x}_1+j(\beta_q^a\mp\beta_p^b)f(\bar{x}_1)\right] \end{matrix}\right\} d\bar{x}_1, \\
		& + \sum\limits_{q=-\infty}^{\infty} a_q^- \frac{1}{\Lambda} \int\limits_0^{\Lambda} \left\{\begin{matrix} \frac{\left(\beta_p^b\mp \frac{\chi_b}{\chi_a}\beta_q^a\right)\mp\left(\frac{\chi_b}{\chi_a}k_{1q}+k_{1p}\right)f'(\bar{x}_1)}{2\beta_p^b} \\ \times\exp\left[j\Delta k_{1qp}\bar{x}_1 - j(\beta_q^a\pm\beta_p^b)f(\bar{x}_1)\right] \end{matrix}\right\} d\bar{x}_1,
	\end{split}
	\label{eq:ab}
\end{equation}
where $\Delta k_{1qp} = k_{1q}-k_{1p}$. The obtained relations between amplitudes $a_p^{\pm}$ and $b_p^{\pm}$ can be rewritten in the $T$-matrix form:
\begin{equation}
	\left(\begin{matrix}b_p^+\\b_p^-\end{matrix}\right) = \sum\limits_{q=-\infty}^{\infty} \left(\begin{matrix}T^{++}_{pq} & T^{+-}_{pq}\\T^{-+}_{pq}&T^{--}_{pq}\end{matrix}\right)\left(\begin{matrix}a_p^+\\a_p^-\end{matrix}\right).
	\label{eq:Tmat_def}
\end{equation}
Therefore, components of the $T$ matrix coming from Eq.~(\ref{eq:ab}) explicitly read
\begin{equation}
	\begin{split}
		&T_{pq}^{++} = \frac{1}{\Lambda} \int\limits_0^{\Lambda} \frac{\left(\beta_p^b + \frac{\chi_b}{\chi_a}\beta_q^a\right) - \left(\frac{\chi_b}{\chi_a}k_{1q}+k_{1p}\right)f'(\bar{x}_1)}{2\beta_p^b} \exp\left[j\Delta k_{1qp}\bar{x}_1+j(\beta_q^a - \beta_p^b)f(\bar{x}_1)\right] d\bar{x}_1, \\
		&T_{pq}^{+-} = \frac{1}{\Lambda} \int\limits_0^{\Lambda} \frac{\left(\beta_p^b - \frac{\chi_b}{\chi_a}\beta_q^a\right) + \left(\frac{\chi_b}{\chi_a}k_{1q}+k_{1p}\right)f'(\bar{x}_1)}{2\beta_p^b} \exp\left[j\Delta k_{1qp}\bar{x}_1+j(\beta_q^a + \beta_p^b)f(\bar{x}_1)\right] d\bar{x}_1, \\
		&T_{pq}^{-+} = \frac{1}{\Lambda} \int\limits_0^{\Lambda} \frac{\left(\beta_p^b - \frac{\chi_b}{\chi_a}\beta_q^a\right) - \left(\frac{\chi_b}{\chi_a}k_{1q}+k_{1p}\right)f'(\bar{x}_1)}{2\beta_p^b} \exp\left[j\Delta k_{1qp}\bar{x}_1 - j(\beta_q^a + \beta_p^b)f(\bar{x}_1)\right] d\bar{x}_1, \\
		&T_{pq}^{--} = \frac{1}{\Lambda} \int\limits_0^{\Lambda} \frac{\left(\beta_p^b + \frac{\chi_b}{\chi_a}\beta_q^a\right) + \left(\frac{\chi_b}{\chi_a}k_{1q}+k_{1p}\right)f'(\bar{x}_1)}{2\beta_p^b} \exp\left[j\Delta k_{1qp}\bar{x}_1 - j(\beta_q^a - \beta_p^b)f(\bar{x}_1)\right] d\bar{x}_1.
	\end{split}
	\label{eq:T1}
\end{equation}
Supposing $\beta_q^a\neq\beta_p^b$, this can be simplified:
\begin{equation}
	\begin{split}
		&T_{pq}^{++} = \zeta^+_{pq}\frac{1}{\Lambda} \int\limits_{0}^{\Lambda} \exp\left[ j(q-p)K\bar{x}_1 + j(\beta_q^a-\beta_p^b)f(\bar{x}_1) \right] d\bar{x}_1,\\
		&T_{pq}^{+-} = \zeta^-_{pq}\frac{1}{\Lambda} \int\limits_{0}^{\Lambda} \exp\left[ j(q-p)K\bar{x}_1 - j(\beta_q^a+\beta_p^b)f(\bar{x}_1) \right] d\bar{x}_1,\\
		&T_{pq}^{-+} = \zeta^-_{pq}\frac{1}{\Lambda} \int\limits_{0}^{\Lambda} \exp\left[ j(q-p)K\bar{x}_1 + j(\beta_q^a+\beta_p^b)f(\bar{x}_1) \right] d\bar{x}_1,\\
		&T_{pq}^{--} = \zeta^+_{pq}\frac{1}{\Lambda} \int\limits_{0}^{\Lambda} \exp\left[ j(q-p)K\bar{x}_1 - j(\beta_q^a-\beta_p^b)f(\bar{x}_1) \right] d\bar{x}_1,
	\end{split}
	\label{eq:T2}
\end{equation}
with $K=2\pi/\Lambda$, and constant factors
\begin{equation}
	\zeta_{pq}^{\pm} = \frac{\pm\omega^2\mu_b(\varepsilon_a-\varepsilon_b) + \left(1-\frac{\mu_b}{\mu_a}\right)(\beta_q^a\beta_p^b\pm k_{1p}k_{1q})}{2\beta_p^b(\beta_q^a\mp\beta_p^b)}
	\label{eq:TC_TE}
\end{equation}
for TE modes, and
\begin{equation}
	\zeta_{pq}^{\pm} = \frac{\pm\omega^2\varepsilon_b(\mu_a-\mu_b) + \left(1-\frac{\varepsilon_b}{\varepsilon_a}\right)(\beta_q^a\beta_p^b\pm k_{1p}k_{1q})}{2\beta_p^b(\beta_q^a\mp\beta_p^b)}
	\label{eq:TC_TM}
\end{equation}
for TM modes.

Thus, the following is claimed: any solution of the electromagnetic problem in the transformed medium can be represented by a linear combination of modal fields. Components of the $T$-matrix relating the modal amplitudes at the interface are expressed analytically.

Concerning application of Eqs.~(\ref{eq:T2}), (\ref{eq:TC_TE}), and (\ref{eq:TC_TM}) to the considered examples, components of $T$-matrix of a sinusoidal grating are
\begin{equation}
	\begin{split}
		&T_{pq}^{++} = \zeta^+_{pq} J_{p-q}\left[\frac{1}{2}(\beta_q^a-\beta_p^b)h\right],\\
		&T_{pq}^{+-} = \zeta^-_{pq} J_{p-q}\left[-\frac{1}{2}(\beta_q^a+\beta_p^b)h\right],\\
		&T_{pq}^{-+} = \zeta^-_{pq} J_{p-q}\left[\frac{1}{2}(\beta_q^a+\beta_p^b)h\right],\\
		&T_{pq}^{--} = \zeta^+_{pq} J_{p-q}\left[-\frac{1}{2}(\beta_q^a-\beta_p^b)h\right].
	\end{split}
	\label{eq:T_sin}
\end{equation}
Components of $T$-matrix of a saw-tooth grating are
\begin{equation}
	\begin{split}
		&T_{pq}^{++} = \zeta^+_{pq}\frac{d_1}{\Lambda} \exp\left[j\pi(q-p)\frac{d_1}{\Lambda}\right] \sinc\left[\pi(q-p)\frac{d_1}{\Lambda} - (\beta_q^a-\beta_p^b)\frac{h}{2} \right] \\
		&\quad + \zeta^+_{pq}\frac{d_2}{\Lambda} \exp\left[-j\pi(q-p)\frac{d_2}{\Lambda}\right] \sinc\left[-\pi(q-p)\frac{d_2}{\Lambda} - (\beta_q^a-\beta_p^b)\frac{h}{2} \right], \\
		&T_{pq}^{+-} = \zeta^-_{pq}\frac{d_1}{\Lambda} \exp\left[j\pi(q-p)\frac{d_1}{\Lambda}\right] \sinc\left[\pi(q-p)\frac{d_1}{\Lambda} + (\beta_q^a+\beta_p^b)\frac{h}{2} \right] \\
		&\quad + \zeta^-_{pq}\frac{d_2}{\Lambda} \exp\left[-j\pi(q-p)\frac{d_2}{\Lambda}\right] \sinc\left[-\pi(q-p)\frac{d_2}{\Lambda} + (\beta_q^a+\beta_p^b)\frac{h}{2} \right], \\
		&T_{pq}^{-+} = \zeta^-_{pq}\frac{d_1}{\Lambda} \exp\left[j\pi(q-p)\frac{d_1}{\Lambda}\right] \sinc\left[\pi(q-p)\frac{d_1}{\Lambda} - (\beta_q^a+\beta_p^b)\frac{h}{2} \right] \\
		&\quad + \zeta^-_{pq}\frac{d_2}{\Lambda} \exp\left[-j\pi(q-p)\frac{d_2}{\Lambda}\right] \sinc\left[-\pi(q-p)\frac{d_2}{\Lambda} - (\beta_q^a+\beta_p^b)\frac{h}{2} \right], \\
		&T_{pq}^{--} = \zeta^+_{pq}\frac{d_1}{\Lambda} \exp\left[j\pi(q-p)\frac{d_1}{\Lambda}\right] \sinc\left[\pi(q-p)\frac{d_1}{\Lambda} + (\beta_q^a-\beta_p^b)\frac{h}{2} \right] \\
		&\quad + \zeta^+_{pq}\frac{d_2}{\Lambda} \exp\left[-j\pi(q-p)\frac{d_2}{\Lambda}\right] \sinc\left[-\pi(q-p)\frac{d_2}{\Lambda} + (\beta_q^a-\beta_p^b)\frac{h}{2} \right].
	\end{split}
	\label{eq:T_saw}
\end{equation}

\section{Equivalence between the Rayleigh hypothesis and the association of the coordinate transformation and the modal method}

In accordance with the modal analysis made in the previous section, the transverse field solution of the transformed problem writes
\begin{equation}
	\bar{F}_2 = \left\{
	\begin{split}
		&\sum\limits_{p=-\infty}^{\infty} b_p^+ \exp \left\{ jk_{1p}\bar{x}_1 + j\beta_p^b[\bar{x}_3 + f(\bar{x}_1)] \right\} + b_p^- \exp \left\{ jk_{1p}\bar{x}_1 - j\beta_p^b[\bar{x}_3 + f(\bar{x}_1)] \right\}, \bar{x}_3\leq 0, \\
		&\sum\limits_{p=-\infty}^{\infty} a_p^+ \exp \left\{ jk_{1p}\bar{x}_1 + j\beta_p^a[\bar{x}_3 + f(\bar{x}_1)] \right\} + a_p^- \exp \left\{ jk_{1p}\bar{x}_1 - j\beta_p^a[\bar{x}_3 + f(\bar{x}_1)] \right\}, \bar{x}_3> 0.
	\end{split}
	\right.
	\label{eq:RayleighZ}
\end{equation}
with the modal amplitudes $a_p^{\pm}$ and $b_p^{\pm}$ related by Eq.~(\ref{eq:ab}). As follows from the claim of the second section, the same expression gives the transverse field solution for the initial diffraction problem in the curvilinear coordinate system, i.e., one can perform substitutions $(\bar{x}_1,\bar{x}_2,\bar{x}_3) \rightarrow (z^1,z^2,z^3)$, and $\bar{F}\rightarrow\tilde{F}$. The dispersion relation of Eq.~(\ref{eq:beta}) remains untouched. Using transformation of Eq.~(\ref{eq:transform_zx}) this solution can be rewritten in the initial Cartesian system as the solution to the initial diffraction problem:
\begin{equation}
	F_2 = \left\{
	\begin{split}
		&\sum\limits_{p=-\infty}^{\infty} b_p^+ \exp \left( jk_{1p}x_1 + j\beta_p^bx_3 \right) + b_p^- \exp \left( jk_{1p}x_1 - j\beta_p^bx_3 \right), x_3\leq f(x_1), \\
		&\sum\limits_{p=-\infty}^{\infty} a_p^+ \exp \left( jk_{1p}x_1 + j\beta_p^ax_3 \right) + a_p^- \exp \left( jk_{1p}x_1 - j\beta_p^ax_3 \right), x_3> f(x_1).
	\end{split}
	\right.
	\label{eq:RayleighX}
\end{equation}
The obtained Eq.~(\ref{eq:RayleighX}) is nothing but the Rayleigh expansion in both media above and below the periodic corrugation interface. Invoking the radiation condition one can see that the validity of the RH comes from the validity of the modal expansion in the transformed medium. In other words, completeness of the Rayleigh expansion is defined by completeness of this modal expansion. Modal expansions in the grating diffraction theory are widely considered and proved in certain cases to yield complete solutions of corresponding boundary value problems (e.g.,  \cite{Botten1981-1,Li1993,Adams2008,Gralak2012}), as well as they demonstrated their numerical validity for gratings of arbitrary depth. Nevertheless, a rigorous mathematical proof is required for the general case presented here, and will be given elsewhere.

\section{Discussion and conclusions}
Conventionally the diffraction problem for gratings of finite depth is solved numerically by various methods. In this work a closed analytic solution to the diffraction problem is found in the form of $T$-matrix Eq.~(\ref{eq:T2}) relating amplitudes of diffraction waves from both sides of the grating corrugation profile. Very often such form of a solution is not sufficient since usually one looks for the scattering $S$-matrix of the grating, which is defined as follows:
\begin{equation}
	\left(\begin{matrix} b_{p}^{+} \\ a_p^- \end{matrix}\right) = \sum\limits_{q=-\infty}^{\infty} \left(\begin{matrix} S_{pq}^{bb} & S_{pq}^{ba} \\ S_{pq}^{ab} & S_{pq}^{aa} \end{matrix}\right) \left(\begin{matrix} b_q^- \\ a_q^+ \end{matrix}\right)
	\label{eq:S-matrix}
\end{equation}
Transformation of a $T$-matrix into the $S$ counterpart is not an easy numerical problem. The difficulty is due to the fast exponential growth of some $T$-matrix elements when difference $|p-q|$ increases. The direct inversion of a truncated T matrix often leads to numerical problems even in cases when the solution for infinite matrices exists. It is worth noting here that similar problems arise when transforming numerically an $S$-matrix of a sufficiently deep grating into a diffraction $T$-matrix. All terms of an $S$-matrix are finite but inverting its truncated part leads to numerical instabilities. A way to overcome this issue can be analogous to the approach of \cite{Tishchenko2009}, i.e. the use of multiple precision arithmetic \cite{Semenikhin2013}.

However, the obtained analytic solution presents several advantages. First, it opens an opportunity to transform the $T$-matrix analytically. Second, it displays the structure of the $T$-matrix assisting in its smart transformation. Third, in some cases (like the inverse problem, for example) the solution in the form of a $T$-matrix can be sufficient.

To conclude, this article supports numerical results and develops ideas presented in \cite{Tishchenko2009,Tishchenko2010}. We have combined together two techniques reputed as rigorous in the grating theory: the coordinate transformation of the CM and the TMM. Such direct and rigorous combination yielded two encouraging results. First, the exact electromagnetic solution to a general diffraction problem is found in a closed analytical form. Second, the validity of the RH and related completeness of the Rayleigh expansion in the grating region are shown to be defined by completeness of the true modal expansion in the transformed medium.

\section*{Funding}
The work was supported in part by the Russian Foundation for Basic Research (16-29-11747-ofi-m).

\section*{Appendix}
In this Appendix we prove integrals of Eqs.~(\ref{eq:normint1}) and (\ref{eq:normint2}). Consider a continuous periodic function consisting of $N$ twice differentiable pieces:
\begin{equation}
	f(\xi) = f_i(\xi),\;\xi_{i-1}\leq\xi<\xi_i,\;i=1,\dots,N
	\label{eq:f_piece}
\end{equation}
with $\xi_0=0$ and $\xi_{N} = \Lambda$ and a periodicity condition $f_1(\xi_0)=f_N(\xi_N)$. Eq.~(\ref{eq:beta}) gives
\begin{equation}
	(k_{1p} + k_{1q})(k_{1p} - k_{1q}) + (\beta_p + \beta_q)(\beta_p - \beta_q) = 0.
	\label{eq:kb_relation}
\end{equation}
Here we omit the upper index of the propagation constants $\beta_p$ as the derivation is the same for both upper and lower parts of the transformed medium. If $p=q$ integration is straightforward:
\begin{equation}
	\begin{split}
	&\delta_{pq}\frac{1}{\Lambda} \int\limits_0^{\Lambda} \left[ (\beta_p + \beta_q)\mp(k_{1p}+k_{1q})f'(\xi) \right] \exp\left[ (k_{1q} - k_{1p})\xi\pm(\beta_q-\beta_p)f(\xi) \right] d\xi \\
	&= 2\beta_p \mp \frac{2}{\Lambda}k_{1p} \sum\limits_{i=1}^N \left[f(\xi_i) - f(\xi_{i-1})\right] = 2\beta_p
	\end{split}
	\label{eq:integration_pp_1}
\end{equation}
\begin{equation}
	\begin{split}
	&\delta_{pq}\frac{1}{\Lambda} \int\limits_0^{\Lambda} \left[ (\beta_p - \beta_q)\pm(k_{1p}+k_{1q})f'(\xi) \right] \exp\left[ (k_{1q} - k_{1p})\xi\pm(\beta_p+\beta_q)f(\xi) \right] d\xi \\
	& = \left\{
		\begin{split}
		&\frac{k_{1p}}{j\beta_p\Lambda} \sum\limits_{i=1}^N \left\{\exp\left[\pm 2j\beta_pf_i(\xi_i)\right] - \exp\left[\pm 2j\beta_pf_i(\xi_{i-1})\right]\right\},\;\beta_p\neq 0 \\
		&\pm 2k_{1p}\left[f_i(\xi_i) - f_i(\xi_{i-1})\right],\;\beta_p=0
		\end{split}
	\right.
	= 0.
	\end{split}
	\label{eq:integration_pp_2}
\end{equation}
If $p\neq q$ we multiply a numerator and a denominator under integrals by $(k_{1q}-k_{1p})$ and use Eq.~(\ref{eq:kb_relation}) to get
\begin{equation}
	\begin{split}
	&\frac{1}{\Lambda} \int\limits_0^{\Lambda} \left[ (\beta_p + \beta_q)\mp(k_{1p}+k_{1q})f'(\xi) \right] \exp\left[ (k_{1q} - k_{1p})\xi\pm(\beta_q-\beta_p)f(\xi) \right] d\xi \\
	&= \sum\limits_{i=1}^N \frac{\beta_p+\beta_q}{k_{1q}-k_{1p}} \frac{1}{\Lambda} \int\limits_{\xi_{i-1}}^{\xi_i} \left[ (\beta_p + \beta_q)\mp(k_{1p}+k_{1q})f_i'(\xi) \right] \exp\left[ (k_{1q} - k_{1p})\xi\pm(\beta_q-\beta_p)f_i(\xi) \right] d\xi \\
	&= \frac{\beta_p+\beta_q}{j(k_{1q}-k_{1p})\Lambda} \sum\limits_{i=1}^N \left\{ \begin{matrix} \exp\left[j(k_{1q}-k_{1p})\xi_i \pm j(\beta_q-\beta_p)f_i(\xi_i)\right] \\ - \exp\left[j(k_{1q}-k_{1p})\xi_{i-1} \pm j(\beta_q-\beta_p)f_i(\xi_{i-1})\right] \end{matrix} \right\} \\
	&= \frac{\beta_p+\beta_q}{j(k_{1q}-k_{1p})\Lambda} \exp\left[\pm j(\beta_q-\beta_p)f_1(0)\right] \left\{ \exp\left[j(k_{1q}-k_{1p})\Lambda\right] -1 \right\} = 0,
	\end{split}
	\label{eq:integration_pq_1}
\end{equation}
\begin{equation}
	\begin{split}
	&\frac{1}{\Lambda} \int\limits_0^{\Lambda} \left[ (\beta_p - \beta_q)\pm(k_{1p}+k_{1q})f'(\xi) \right] \exp\left[ (k_{1q} - k_{1p})\xi\pm(\beta_p+\beta_q)f(\xi) \right] d\xi \\
	&= \sum\limits_{i=1}^N \frac{\beta_p-\beta_q}{k_{1q}-k_{1p}} \frac{1}{\Lambda} \int\limits_{\xi_{i-1}}^{\xi_i} \left[ (k_{1q} - k_{1p})\pm(\beta_p+\beta_q)f_i'(\xi) \right] \exp\left[ (k_{1q} - k_{1p})\xi\pm(\beta_q+\beta_p)f_i(\xi) \right] d\xi \\
	&= \frac{\beta_p-\beta_q}{j(k_{1q}-k_{1p})\Lambda} \sum\limits_{i=1}^N \left\{ \begin{matrix} \exp\left[j(k_{1q}-k_{1p})\xi_i \pm j(\beta_q+\beta_p)f_i(\xi_i)\right] \\ - \exp\left[j(k_{1q}-k_{1p})\xi_{i-1} \pm j(\beta_q+\beta_p)f_i(\xi_{i-1})\right] \end{matrix} \right\} \\
	&= \frac{\beta_p-\beta_q}{j(k_{1q}-k_{1p})\Lambda} \exp\left[\pm j(\beta_q+\beta_p)f_1(0)\right] \left\{ \exp\left[j(k_{1q}-k_{1p})\Lambda\right] -1 \right\} = 0.
	\end{split}
	\label{eq:integration_pq_2}
\end{equation}
This proves Eqs.~(\ref{eq:normint1}) and (\ref{eq:normint2}).

\section*{Acknowledgments}
Despite premature decease in August 2016 Prof. Alexandre V. Tishchenko is stated as the first author of this work due to his major input to the presented results.

\end{document}